\documentclass{jpsj-suppl}

\newcommand{\ee}{e^{+}e^{-}}
\newcommand{\leplep}{\ell^{+}\ell^{-}}

\newcommand{\psip}{\psi^{\prime}}

\newcommand{\pim}{\pi^{-}}
\newcommand{\pip}{\pi^{+}}
\newcommand{\piz}{\pi^{0}}
\newcommand{\ppbar}{p\bar{p}}
\newcommand{\nnbar}{n\bar{n}}
\newcommand{\NNbar}{N\bar{N}}
\newcommand{\pb}{\bar{p}}
\newcommand{\nb}{\bar{n}}

\newcommand{\ccbar}{c\bar{c}}

\newcommand{\BBbar}{B\bar{B}}

\newcommand{\qqbar}{q\bar{q}}

\newcommand{\QQbar}{Q\bar{Q}}

\newcommand{\rt}{\rightarrow}

\newcommand{\etal}{\em et al.}
\newcommand{\jpsi}{J/\psi}

\newcommand{\lm}{\Lambda}

\newcommand{\lmb}{\bar{\Lambda}}

\title{Multiquark Hadrons}

\author{Stephen Lars \textsc{Olsen}$^{1}$}

\inst{$^{1}$ Center for Underground Physics, Institute for Basic Science,
Yuseong-gu, Daejeon 34047, Korea}

\email{solsensnu@gmail.com}

\recdate{December 14, 2015}

\abst{
A number of candidate multiquark hadrons, i.e., particle resonances with substructures that are more complex
than the quark-antiquark mesons and three-quark baryons that are prescribed in the textbooks, have recently
been observed. In this talk I present: some recent preliminary BESIII results on the near-threshold behavior
of $\sigma(\ee\rt\lm\lmb)$ that may or may not be related to multiquark mesons in the light- and strange-quark
sectors; results from Belle and LHCb on the electrically charged, charmoniumlike $Z(4430)^{\pm}\rt\pi^{\pm}\psip$
resonance that necessarily has a four-quark substructure; and the very recent LHCb discovery of the
$P_c(4380)$ and $P_c(4450)$ hidden-charm resonances seen as a complex structure in the $\jpsi\ p$ invariant
mass distribution for $\lm_b\rt K^{-} \jpsi\ p$ decays and necessarily have a five-quark substructure and are,
therefore, prominent candidates for pentaquark baryons.}

\kword{baryon form factors, $XYZ$-mesons, tetraquarks, pentaquarks}

\begin{document}
\maketitle

\section{Introduction}

Gell-Mann, in his original quark model paper, speculated that in addition to the quark-antiquark ($\qqbar$)
mesons and quark-quark-quark ($qqq$) baryons that account for the octets and decuplets of flavor-$SU(3)$,
there could be more complex structures such as $\qqbar\qqbar$ tetraquark mesons and $qqq\qqbar$ pentaquark
baryons~\cite{gellmann64}. Zweig made similar speculations in his contemporaneous paper on
``aces''~\cite{zweig64}.  However, despite a considerable multi-decade experimental effort at searching
for examples of resonances with these multiquark configurations in the light ($u,d$) and strange ($s$)
quark sectors, no compelling multiquark candidates were identified.  On the other hand, during
the last decade a number of electrically charged mesons that contain a heavy quark pair (i.e.,
$\QQbar$, where $Q=c$ or $b$) plus a pair of light quarks have been observed.  (For recent reviews see
refs.~\cite{nora_qwg,bodwin13,olsen15}.)  This year, the LHCb experiment presented striking evidence for
$\jpsi\ p$ resonances in $\lm_b\rt K^- \jpsi\ p$ decays~\cite{lhcb_5q} that are strong candidates for the
pentaquark states that were first suggested by Gell-Mann and Zweig over fifty years ago.

In contrast, there is still
no unambiguous example of a non-standard hadron in the $u,d,s$ quark sector.  Candidates for an $\eta\pi$
meson resonance with an exotic $J^{PC}=1^{-+}$ quantum number assignment are discussed in ref.~\cite{dzierba03};
the interpretation of the anomalously narrow $a_1(1420)$ axial vector state, reported by the COMPASS
experiment~\cite{compass_a1}, as a four-quark state is still not settled.  Anomalies in the light scalar
octet, especially the masses and properties of the $f_0(980)$ and $a_0(980)$ mesons, have been interpreted as
evidence for multi-quark states. Recent measurements by BES~\cite{bes_sigma} and Belle~\cite{belle_f0} have
provided experimental support for this point of view~\cite{achasov_2009}.    In addition, there
are a number of puzzles associated with near-threshold nucleon-antinucleon systems that may
point to six-quark ``baryonium'' $\NNbar$ bound states.  These include a prominent
threshold $\ppbar$ mass peak in radiative $\jpsi\rt\gamma\ppbar$ decays reported by BESII~\cite{bes2_gppb},
hints of anomalous behavior of the time-like neutron form-factor seen in $\ee\rt \nnbar$ by Fenice~\cite{fenice}
and SND~\cite{snd_nnbar}, and signs of an anomalous near-threshold angular distribution for $\ee\rt\ppbar$ 
by BaBar~\cite{babar_ppb} and CMD-3~\cite{cmd_ppb}.

In this talk I review results from near-threshold baryon pair production measurements in $\ee$ annihilations that
hint at a non-analytic behavior of their time-like form factors, including some preliminary results from BESIII
on $\ee\rt\lm\lmb$. In addition, I review the experimental status of the $Z(4430)$ candidate for a tetraquark
meson state and the recent LHCb report on the observation of candidate pentaquark baryons with hidden charm.

\section{Time-like baryon form factors}
\label{sec:BBff}
The Born differential cross section for exclusive spin-1/2 baryon pair ($\BBbar$) production in $\ee$
annihilation is usually expressed in terms of the electric and magnetic electromagnetic form factors,
$G_E$ and $G_M$, as:
\begin{equation}
\frac{d\sigma}{d\Omega}=\frac{\alpha^2\beta {\mathcal C}}{4M^2_{\BBbar}}\left[(1+\cos^2\theta)|G_M(M_{\BBbar})|^2
+\frac{1}{\tau}\sin^2\theta|G_E(M_{\BBbar})|^2\right], 
\label{eqn:BBff}
\end{equation} 
where $M_{\BBbar}~(=\sqrt{s})$ is the $\BBbar$ invariant mass, $\tau=M_{\BBbar}^2/4m_B^2$,
$\beta$ is the center of mass (c.m.) velocity of each $B$, and $\theta$ is the polar angle of the baryons. The factor
${\mathcal C}$ accounts for the Coulomb attraction between charged baryon-antibaryon pairs; in
an approximation where the baryons are point-like, it has the values ${\mathcal C}=1$ for neutral
baryons and ${\mathcal C}=(\pi\alpha /\beta)/[1-\exp(-\pi\alpha /\beta)]$ for charged baryons.
If the form factors are analytic functions, $G_M(2m_B)=G_E(2m_B)$ and the angular distribution
at $M_{\BBbar}^2=4m_B^2$ is isotropic.  Most experiments report cross section values integrated over
$\cos\theta$ in terms of an effective form factor defined as
\begin{equation}
|F_{\rm eff}(M_{\BBbar})|^2=(|G_M(M_{\BBbar})|^2 + \frac{1}{2\tau}|G_E(M_{\BBbar})|^2)/(1+\frac{1}{2\tau}).
\label{eqn:BBffeff}
\end{equation}
In the case of the point-like approximation for ${\mathcal C}$, the charged baryons have a non-zero
cross section at threshold of $\sigma(2m_B) = \pi^2\alpha^3 |G_M|^2/2m_B^2$, and a threshold cross
section for neutral baryons that is zero and grows as $\sigma\propto \beta$ at higher energies.

\begin{figure}[htb]
\begin{minipage}[t]{75mm}
  \includegraphics[height=0.73\textwidth,width=0.95\textwidth]{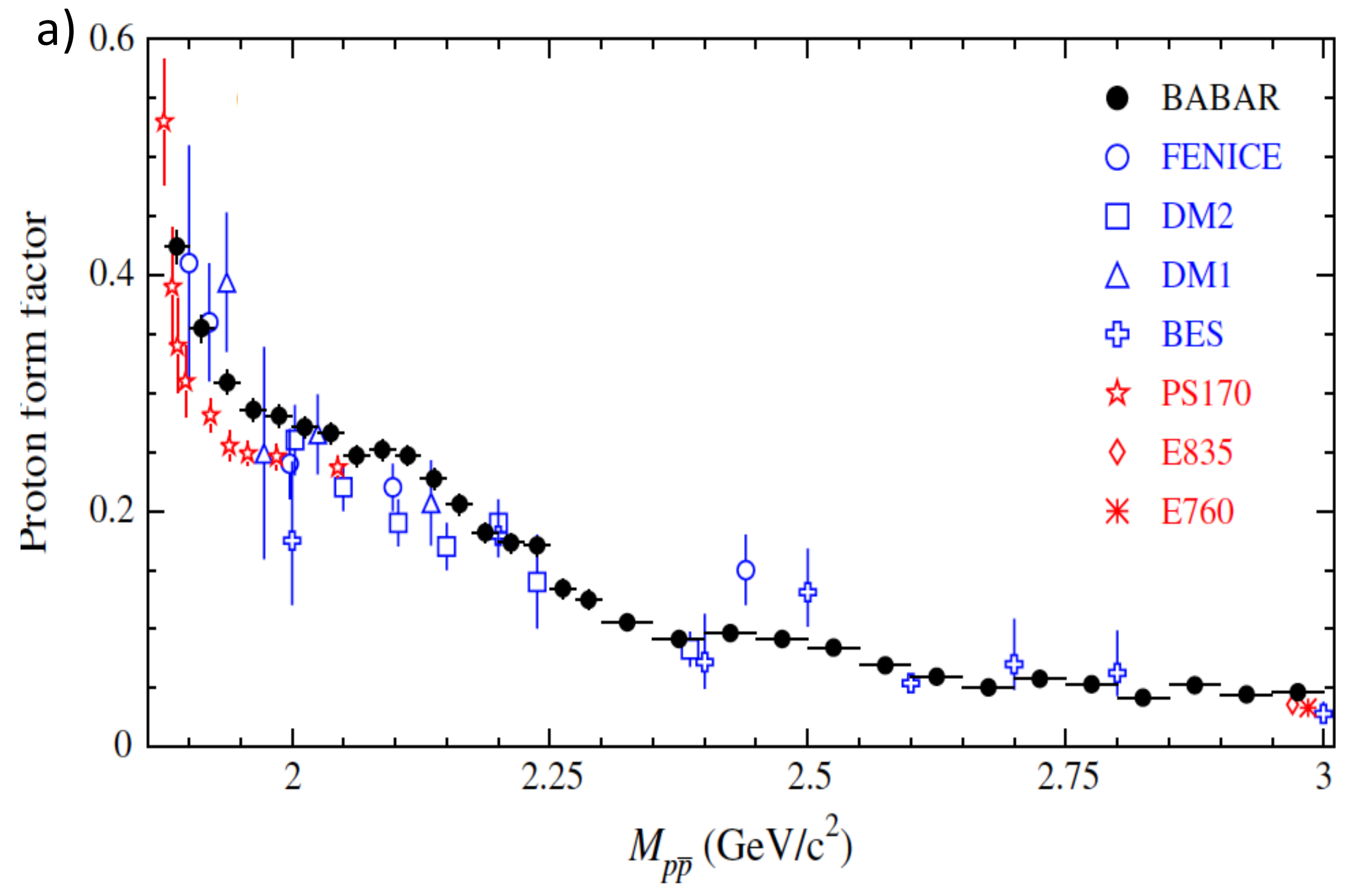}
\end{minipage}
\begin{minipage}[t]{75mm}
  \includegraphics[height=0.75\textwidth,width=0.95\textwidth]{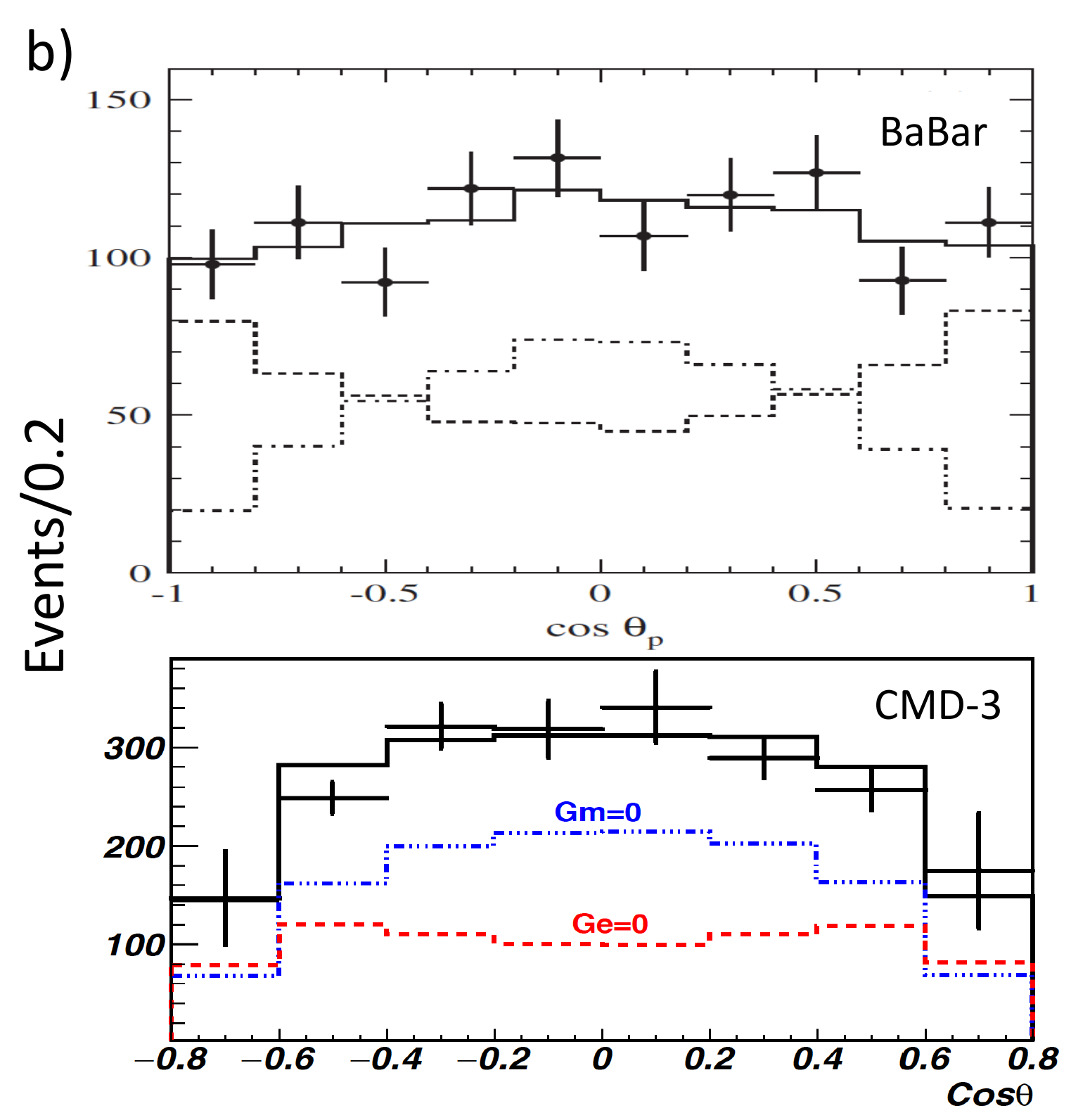}
\end{minipage}
\hspace{\fill}
\caption{\footnotesize {\bf a)} 
(Figure~16a from ref.~\cite{babar_ppb}.)
The effective proton form factor ($|F_{\rm eff}|$) from different experiments.
The measurements from PS 170 are the red stars and those from BaBar are the solid circles. 
{\bf b)}  Near-threshold angular distributions from BaBar {\it (upper)} and CMD-3 {\it (lower)}. 
The dashed (dash-dot) histogram shows the contribution from the $|G_M|$ ($|G_E|$) term in
Eq.~\ref{eqn:BBff}. 
}
\label{fig:NNbar}
\end{figure}

The PS 170 experiment at LEAR used $\ppbar\rt\ee$ reactions to study the proton form-factors and
found a sharp increase as $\beta\rt 0$~\cite{ps170_ppb}.   This rise at small values of $\beta$ was
confirmed by  BaBar measurements for $\beta$ values as low as $\langle \beta \rangle=0.2$ using the
initial state radiation process $\ee\rt\gamma_{\rm isr}\ppbar$ at $\sqrt{s}\simeq 10.6$~GeV~\cite{babar_ppb};
their reported cross section at  $\langle\beta\rangle=0.2$ is $\sigma(\ee\rt\ppbar)=0.53\pm 0.1$~nb (see
Fig.~\ref{fig:NNbar}a). In refs.~\cite{heidenbauer06} and~\cite{gychen10} this enhancement is attributed to
a strong $S$-wave final state interaction (fsi), which effects only the isotropic part of the angular
distribution.  However, somewhat surprisingly, BaBar found evidence for a non-isotropic angular distribution
for events with  $\langle\beta \rangle = 0.2$ that corresponds to  $|G_E/G_M|=1.36\pm 0.15$, a
$\sim 2\sigma$ discrepancy from unity (see the upper panel in Fig.~\ref{fig:NNbar}b). The CMD-3 group's
measurement closest to threshold, at $\langle \beta\rangle =0.11$, is even larger, $|G_E/G_M|=1.49\pm 0.23$,
but also just a $\sim 2\sigma$ deviation from expectations~\cite{cmd_ppb}.

\begin{figure}[htb]
\begin{minipage}[t]{75mm}
  \includegraphics[height=0.73\textwidth,width=0.95\textwidth]{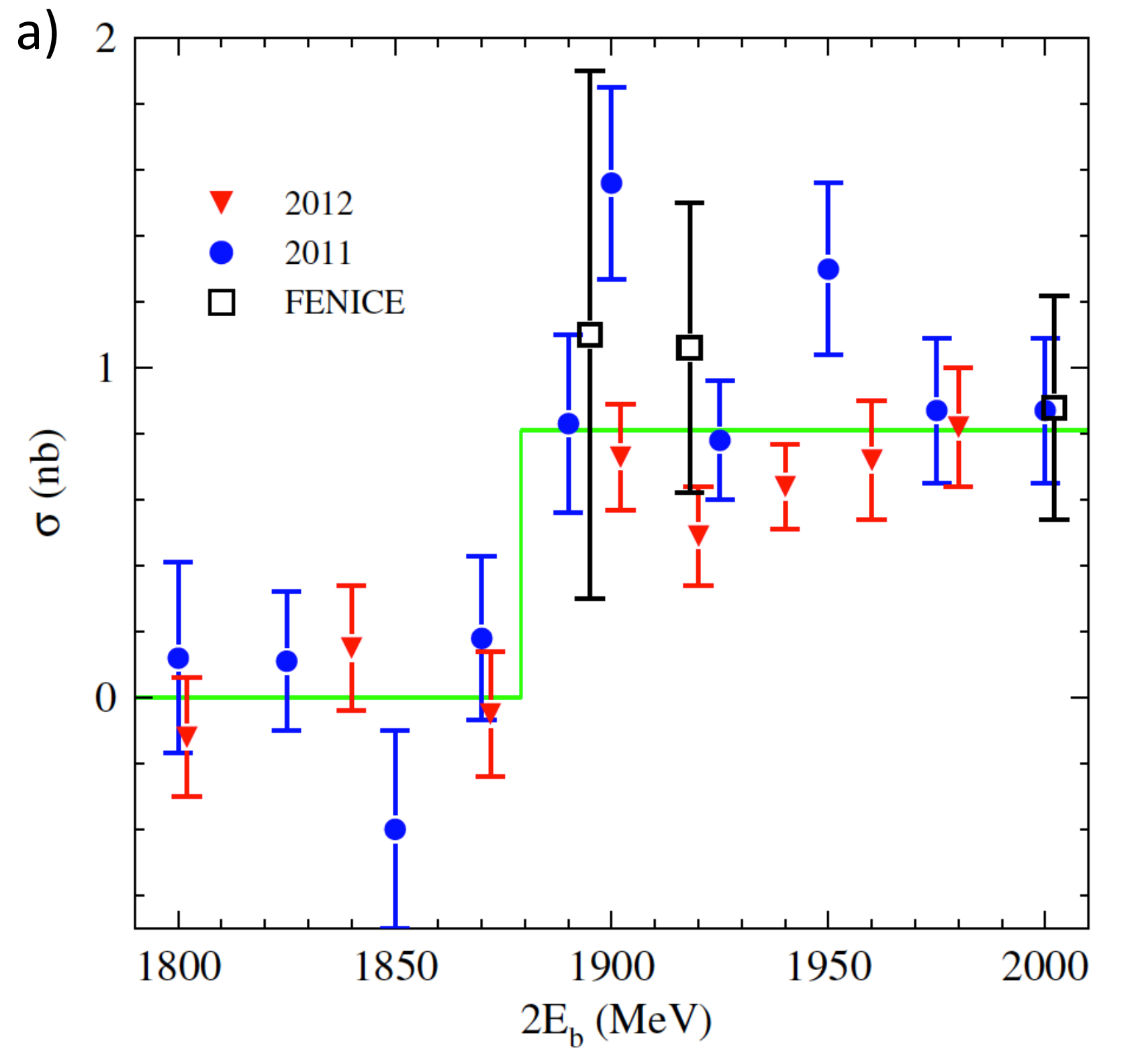}
\end{minipage}
\begin{minipage}[t]{75mm}
  \includegraphics[height=0.73\textwidth,width=0.95\textwidth]{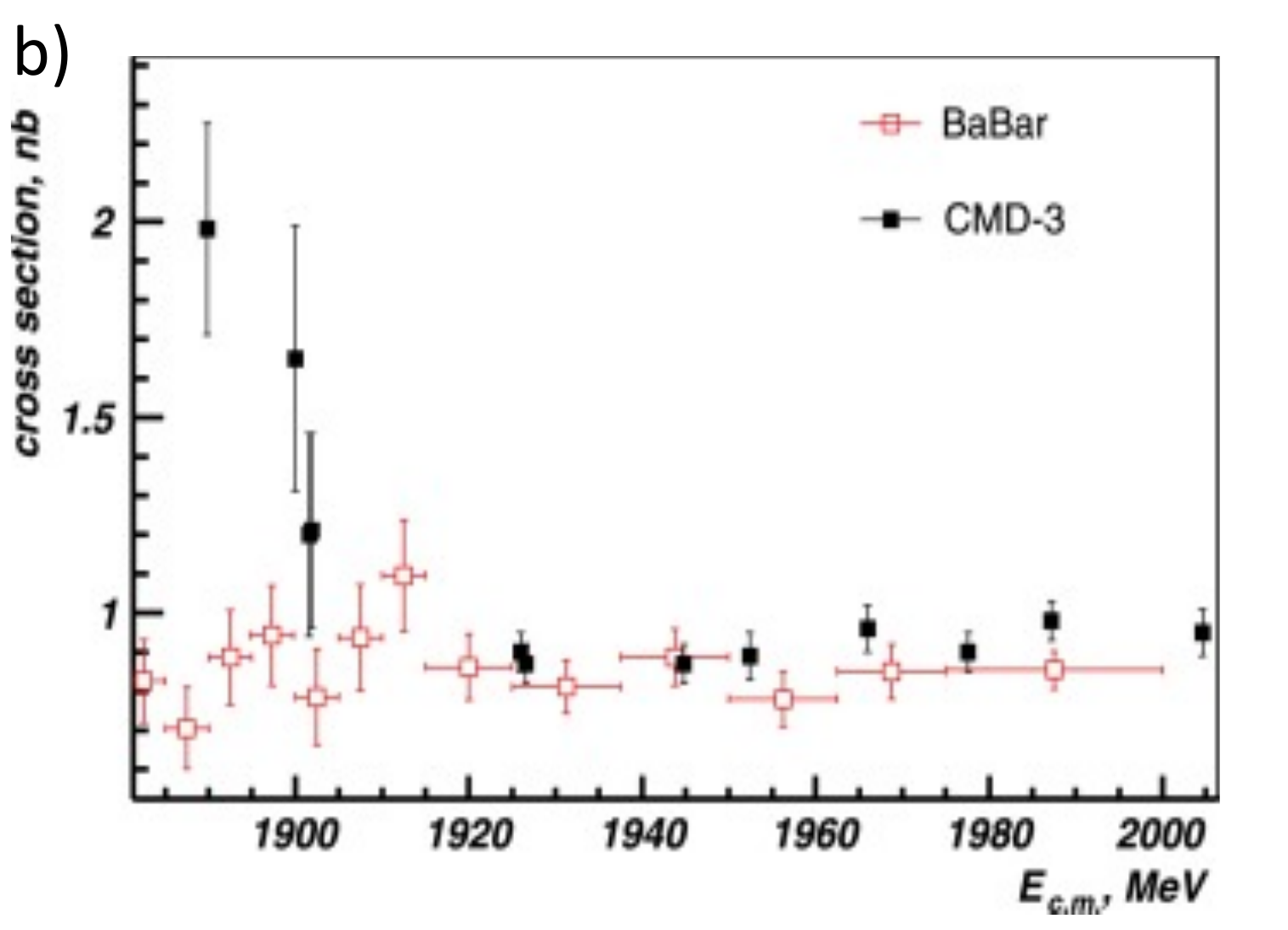}
\end{minipage}
\hspace{\fill}
\caption{\footnotesize {\bf a)}
(Figure~9 from ref.~\cite{snd_nnbar}.)
Measurements of $\sigma(\ee\rt\nnbar)$ from SND (red triangles and solid blue circles) and Fenice
(open black squares). 
{\bf b)}(Figure~16 from ref.~\cite{cmd_ppb}.)
Measurements of $\sigma(\ee\rt\ppbar)$ from BaBar (red inverted triangles) and CMD-3
(solid black squares). No sign of a $\sigma\propto \beta$  behavior at threshold is evident. 
}
\label{fig:binp}
\end{figure}

For $\ee\rt\nnbar$, ${\mathcal C}=1$ and the cross-section is expected to be zero at threshold and grow
as $\sigma(\ee\rt\nnbar)\propto \beta$.  However, there is no sign for this type of behaviour in the
cross section results from SND~\cite{snd_nnbar} and Fenice~\cite{fenice} shown in Fig.~\ref{fig:binp}a. 
The SND cross section value at $\langle \beta\rangle=0.11$ is $0.83\pm 0.27$~nb, which is consistent with
BaBar's mesurements for $\ee\rt\ppbar$, shown in Fig.~\ref{fig:binp}b, even though the Coulomb factor
is expected to be quite different for the two processes.  The CMD-3 experiment's cross section results
are shown in the same figure. While they agree well with BaBar's measurements for c.m. energies above
$\sim 1900$~MeV, their $\sigma(\ee\rt p\bar{p})$ measurements at energies closer to threshold increase
sharply, in contrast to the nearly flat Babar results.   

These are difficult measurements and the experimental errors are large. The point-like approximation
used in the evaluation of of eq.~(\ref{eqn:BBff}) may not be valid. The non-isotropic angular
distributions are measured at $\langle \beta \rangle =0.11$, and may become isotropic at even smaller
$\beta$ values.  On the other hand, the data also point to the possibility that something interesting
may be lurking near the $\sqrt{s}=2m_B$ thresholds. This has stimulated experimental studies at other
baryon thresholds by BESIII, which has recently released preliminary results of cross section
measurements near the $\lm\lmb$ threshold.

\subsection{The near-threshold ${\ee\rt\lm\lmb}$ cross section from BESIII}

\begin{figure}[htb]
\begin{minipage}[t]{75mm}
  \includegraphics[height=0.77\textwidth,width=1.0\textwidth]{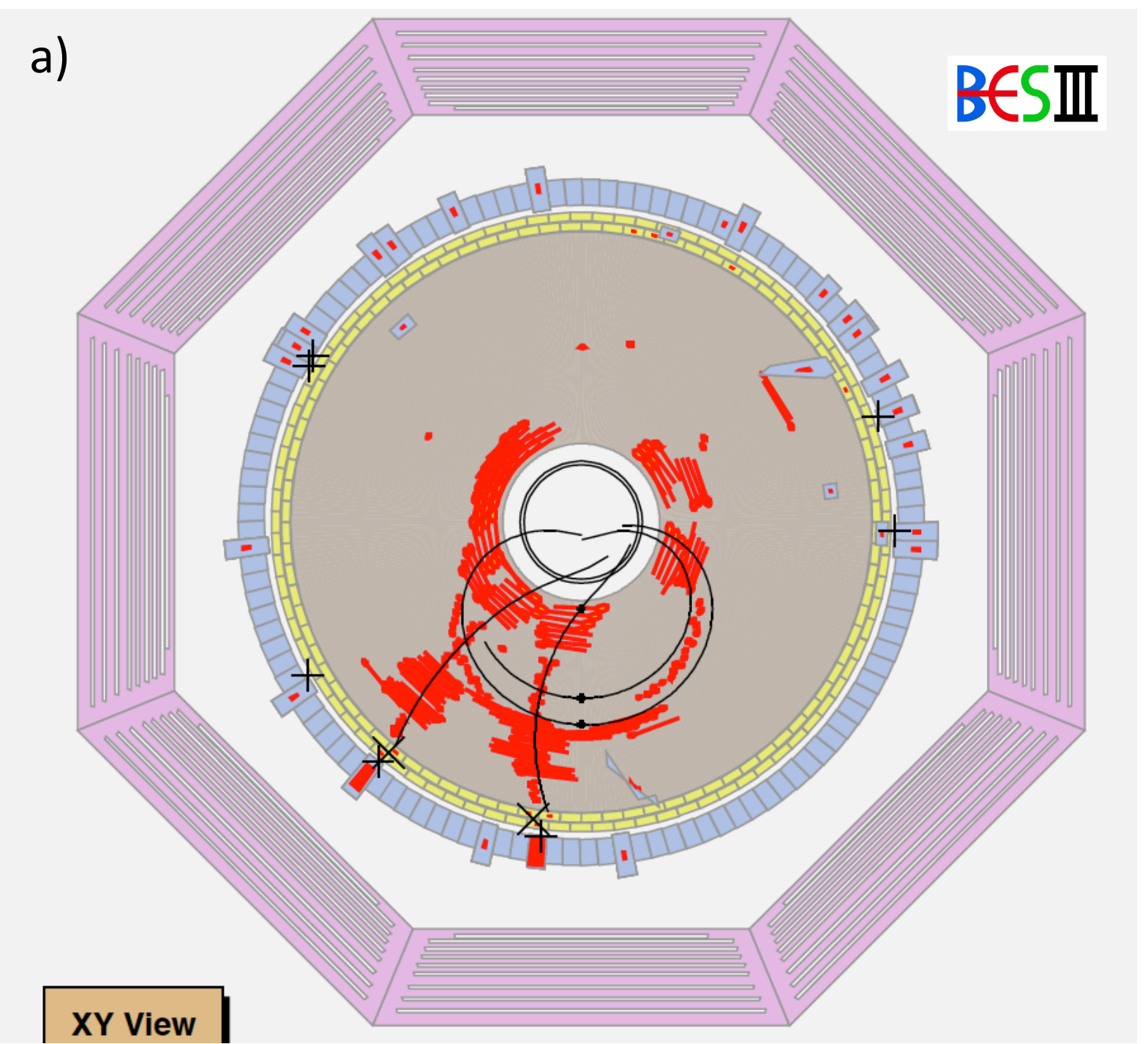}
\end{minipage}
\begin{minipage}[t]{75mm}
  \includegraphics[height=0.77\textwidth,width=1.0\textwidth]{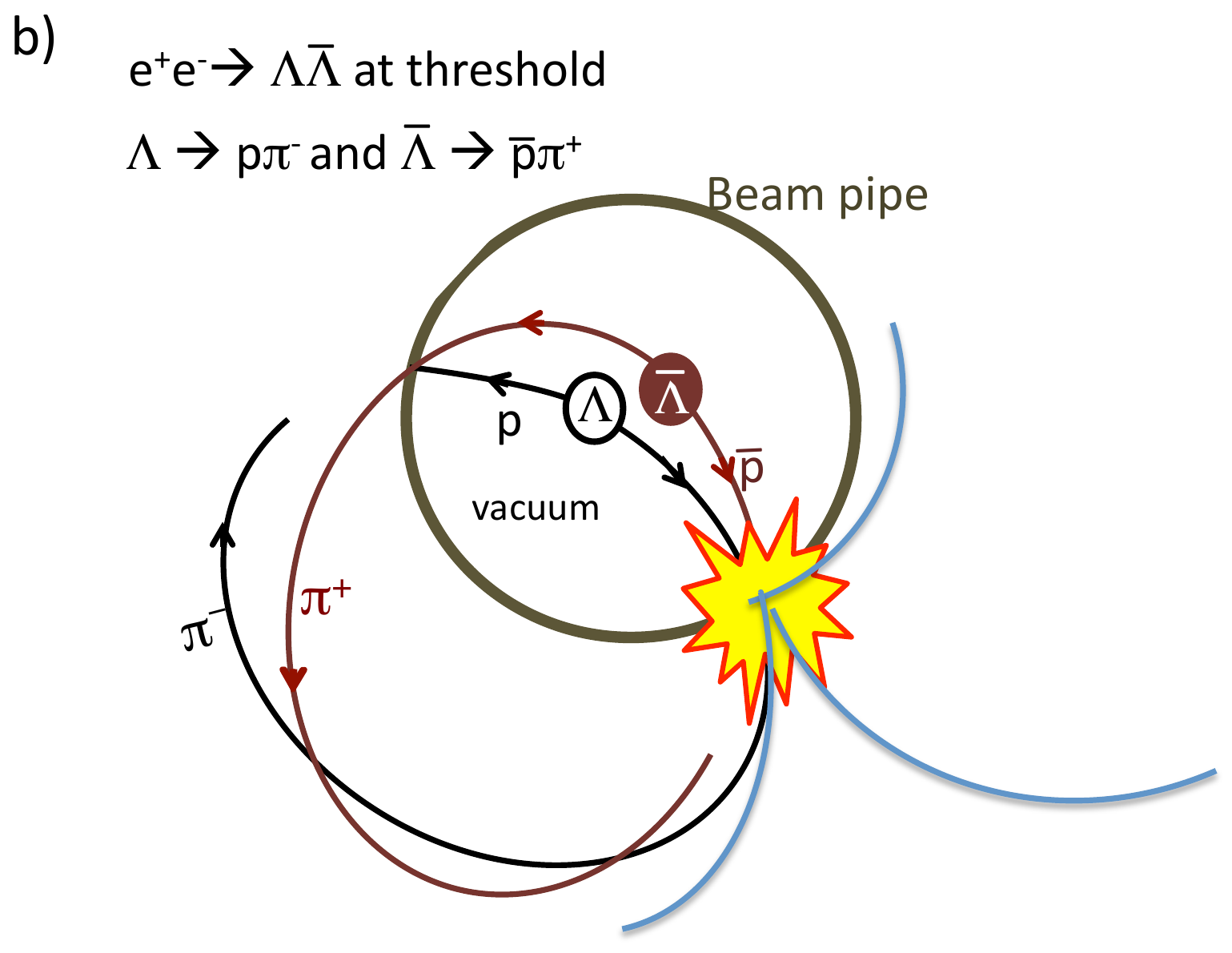}
\end{minipage}
\hspace{\fill}
\caption{\footnotesize {\bf a)} An event in the BESIII detector at $\sqrt{s}=2232.4$~MeV, 1~Mev above
the $\sqrt{s}=2m_{\lm}$ threshold.  
{\bf b)} A cartoon of an $\ee\rt\lm\lmb$ event produced 1~MeV above threshold in which $\lm\rt p\pim$
and $\lmb\rt\pb\pip$.
}
\label{fig:eellb}
\end{figure}

Figure~\ref{fig:eellb}a shows an event in the BESIII detector that is typical of most of the events
in the raw data sample and atypical of the events that are usually shown in a talk like this. 
This particular event has two high momentum tracks that probably triggered the event, but these do
not originate from the $\ee$ interaction point (IP) in the center of the beam pipe, instead they appear
to come  from an interaction in the material of the beam pipe and the inner wall of the detector,
which has an inner radius of 3~cm. A large
fraction of BESIII triggers are due to high momentum tracks produced by electrons (or positrons) in
the BEPCII beams that lose energy via bremsstrahlung on residual gas atoms in the collider vacuum
system or get lost due to the Touchek effect and are over-focused into the beampipe, where they interact.
These account for most of the raw-data events collected by $\ee$ collider experiments, which deliberately
use very loose trigger requirements.  These ``beam-wall'' events are usually filtered out by sophisticated
software during off-line data processing. 

However, the event shown in Fig.~\ref{fig:eellb}a was collected at a c.m. energy that
is only 1~MeV above the $\lm\lmb$ threshold (at $\sqrt{s}=2232.4$~MeV).
In $\ee\rt\lm\lmb$ events at this energy the $\lm$ and $\lmb$ are produced with velocities of $0.03c$
and mean decay distances of 2.4~mm.  The main $\lm$ ($\lmb$) decay modes are $p\pim$
($\pb\pip$) and $n\piz$ ($\nb\piz$); at $\sqrt{s}=2232.4$~MeV the final state pions have
a c.m. three-momentum of $p\simeq 100$~MeV/$c$.

An event where $\lm\rt p\pim$ and $\lmb\rt\pb\pip$, which occurs for 41\% of $\lm\lmb$ events, is
illustrated in the cartoon drawing of Fig.~\ref{fig:eellb}b. The $\pim$ from $\lm\rt p\pim$ will traverse
the inner detector while bending in the detector's 1~T magnetic field, producing a curling helical track
with an origin near the IP and a projected radius $\le$33~cm; the proton, which has a kinetic energy of
only 4.5~MeV, will range out and stop in the material of the beryllium vacuum pipe, where it will spend
eternity.  The $\pip$ from
$\lmb\rt\pb\pip$ will produce a curling track that is similar to that of the $\pim$, except with opposite
sign, and the antiproton will stop in the beam pipe where it will promptly annihilate and typically produce
some higher momentum charged tracks that originate from the $\pb$'s stopping point.  These events are
selected by requiring two oppositely charged pion tracks originating from near the IP, both with
$p\simeq 100$~MeV/$c$, plus at least one higher momentum track that originates from a point $\simeq 3$~cm
radially outside of the IP.  The event shown in Fig.~\ref{fig:eellb}a
satisfies these requirements. 
 
In 36\% of the remaining $\lm\lmb$ events, the $\lmb$ decays to an $\nb\piz$ final state.  In most of
these events the $\nb$ passes through the inner detector and tracking chamber and annihilates in one of
the CsI(Tl) crystals that comprise the electromagnetic calorimeter that surrounds the tracking volume.
The annihilation products produce large energy deposits that are distributed across a number of nearby
crystals in an irregular pattern that is distinct from those produced by high energy $\gamma$-rays.
Accompanying this $\nb$ shower would be two $\gamma$ rays from the decay of the $\piz$ from the $\lmb$
decay, or from the accompanying $\lm$ if it decays to $ n\piz$. These shower pairs have an invariant
mass equal to $m_{\piz}$ and a net laboratory momentum of $105$~MeV/$c$.  

\begin{figure}[htb]
\begin{minipage}[t]{75mm}
  \includegraphics[height=0.73\textwidth,width=0.95\textwidth]{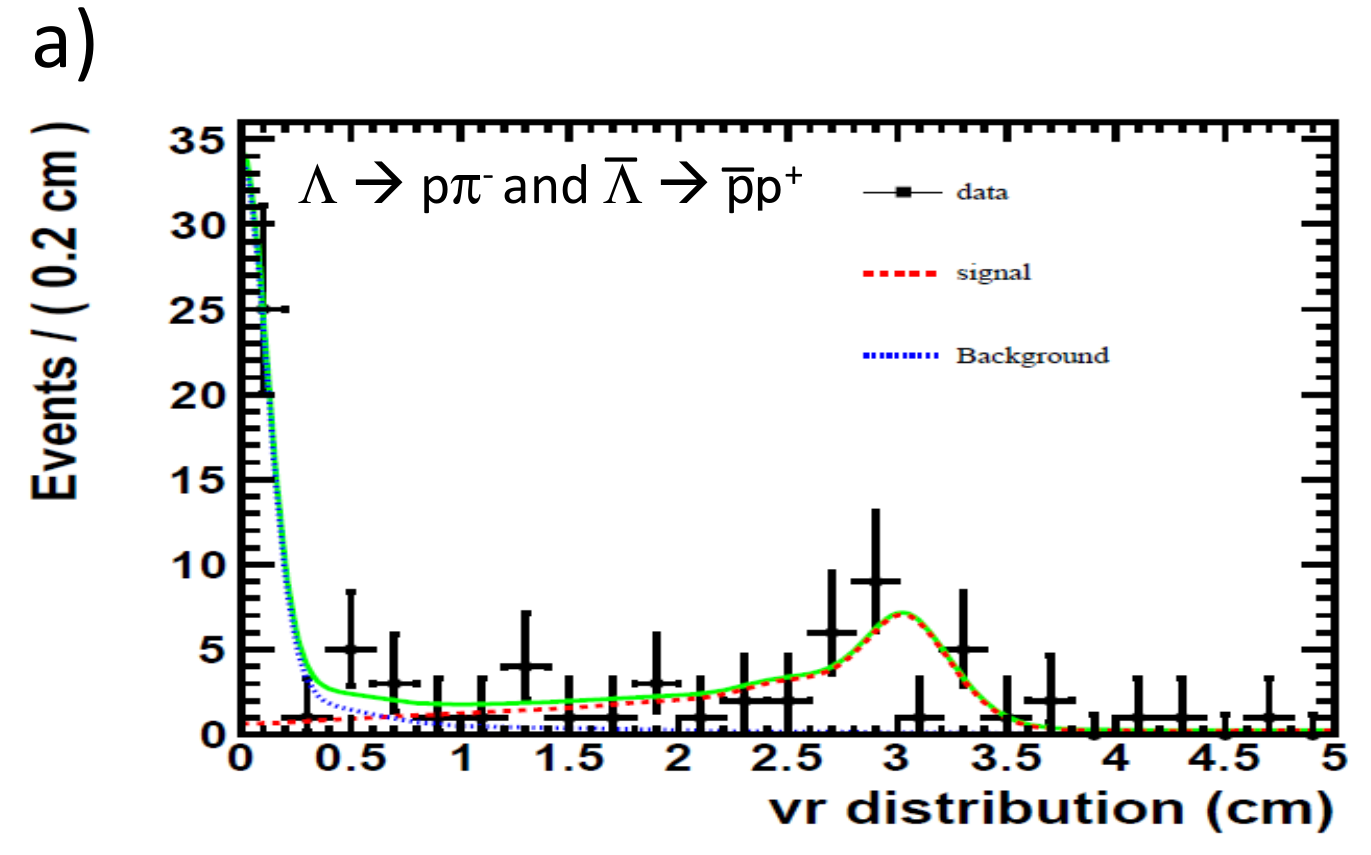}
\end{minipage}
\begin{minipage}[t]{75mm}
  \includegraphics[height=0.73\textwidth,width=0.95\textwidth]{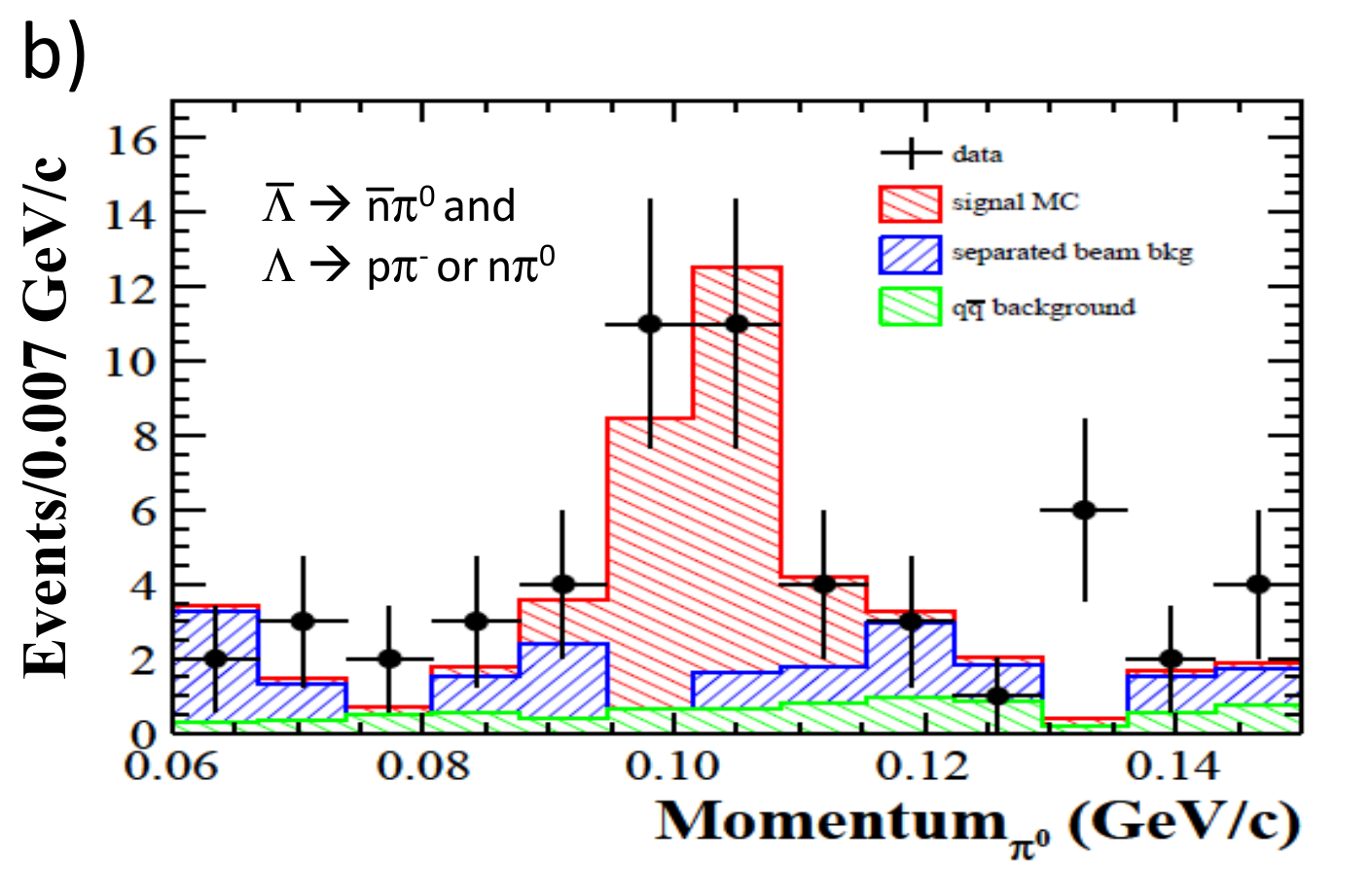}
\end{minipage}
\hspace{\fill}
\caption{\footnotesize {\bf a)}
The $v_r$ distribution for events with a $p\simeq 100$~MeV/$c$ $\pip$ and $\pim$
(BECIII-preliminary).
{\bf b)}
 The momentum distribution of $\piz\rt\gamma\gamma$ candidates
recoiling against an $\nb$-like shower in the BESIII electromagnetic calorimeter.}
\label{fig:eellb-sig}
\end{figure}

Figure~\ref{fig:eellb-sig}a shows the distribution of maximum impact parameter values ($V_r$) for tracks
that accompany two oppositely charged pions with momentum near 100~MeV/$c$ that charcterize
$\lm\rt p\pim$-$\lmb\rt\bar{p}\pip$ events. The peak near $V_r\simeq 3$~cm is a signal expected for $\lm\lmb$
events where both $\lm$s decay to charged final states.  The background,
estimated from charged pion momentum sidebands and data taken with the collider beams separated
at the IP, is shown as a dotted blue curve.  A fit to this distribution gives a $\lm\lmb$ signal yield
of $43\pm 7$ events that translates into a radiatively corrected, Born cross section for $ee\rt\lm\lmb$
of $0.33\pm 0.07$~nb, where here, and elsewhere in this report, the (nearly equal) statistical and
systematic errors are added in quadrature.

The BESIII group made extensive MC studies of potential backgrounds to the $\nb \piz$ signal. The
dominant background from $\ee$ collisions was found to come from $\ee\rt\qqbar$ processes.  A neural-net
event filter was developed to reduce these.  Potential backgrounds from single beam interactions were studied
using separated beam data.   Neither source produces $\piz$'s that make a peak near $p_{\piz}\simeq 105$~MeV/$c$. 
Figure~\ref{fig:eellb-sig}b shows the momentum distribution for $\piz\rt\gamma\gamma$ candidates
in events with an energetic shower in the electromagnetic calorimeter that satisfies the $\nb$ shower selection
requirements (and no more than two accompanying charged tracks). Here a distinct peak centered at
$p\simeq 105$~MeV/$c$ is evident; a fit to this distribution finds a signal yield of $21.8\pm 6.4$ events
that translate into a radiatively corrected Born cross section of
$\sigma(\ee\rt\lm\lmb)=0.30\pm 0.10$~nb, in good agreement with the result from the all-charged-mode result.

\begin{figure}[htb]
\begin{minipage}[t]{75mm}
  \includegraphics[height=0.73\textwidth,width=0.95\textwidth]{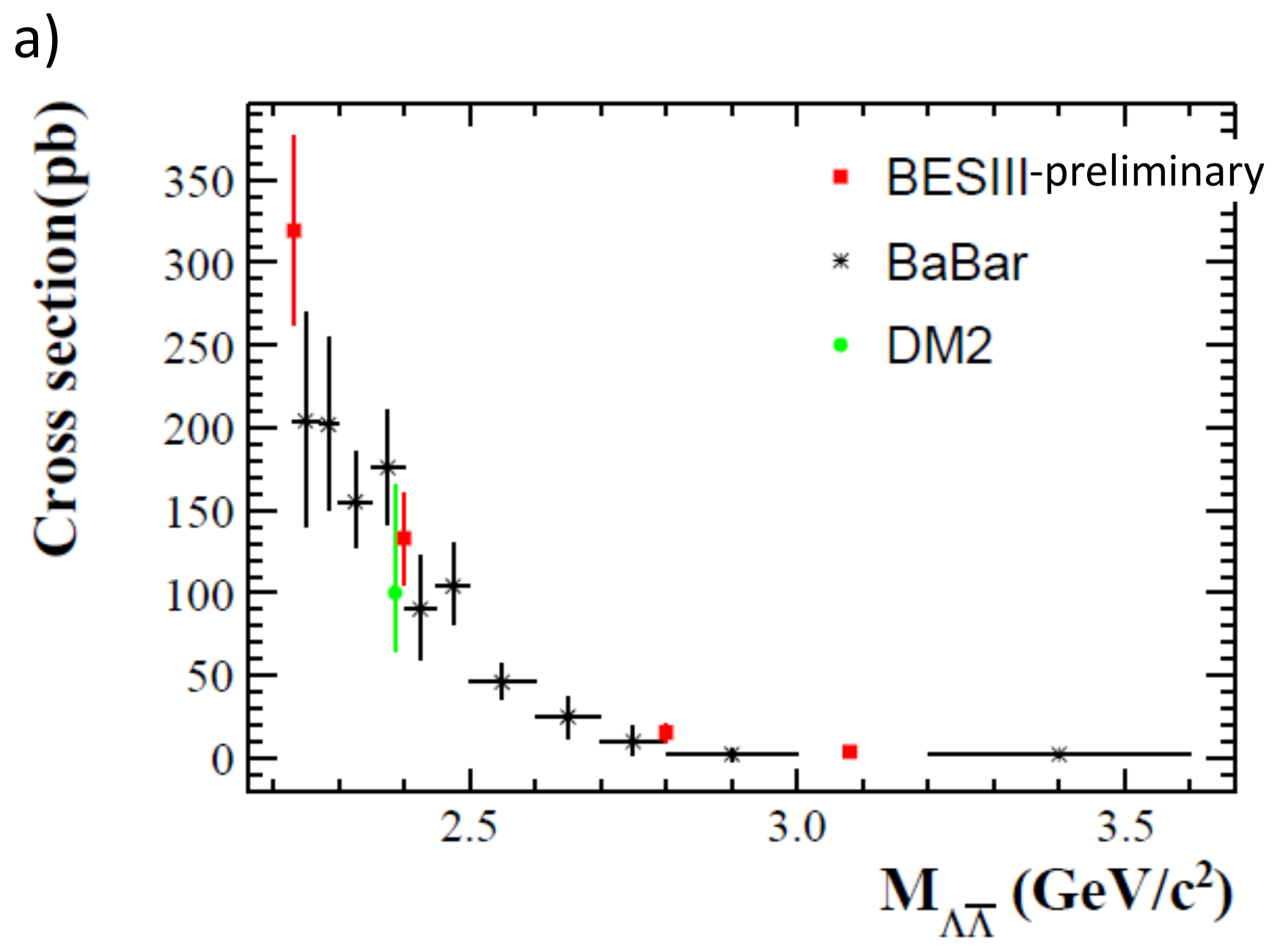}
\end{minipage}
\begin{minipage}[t]{75mm}
  \includegraphics[height=0.73\textwidth,width=0.95\textwidth]{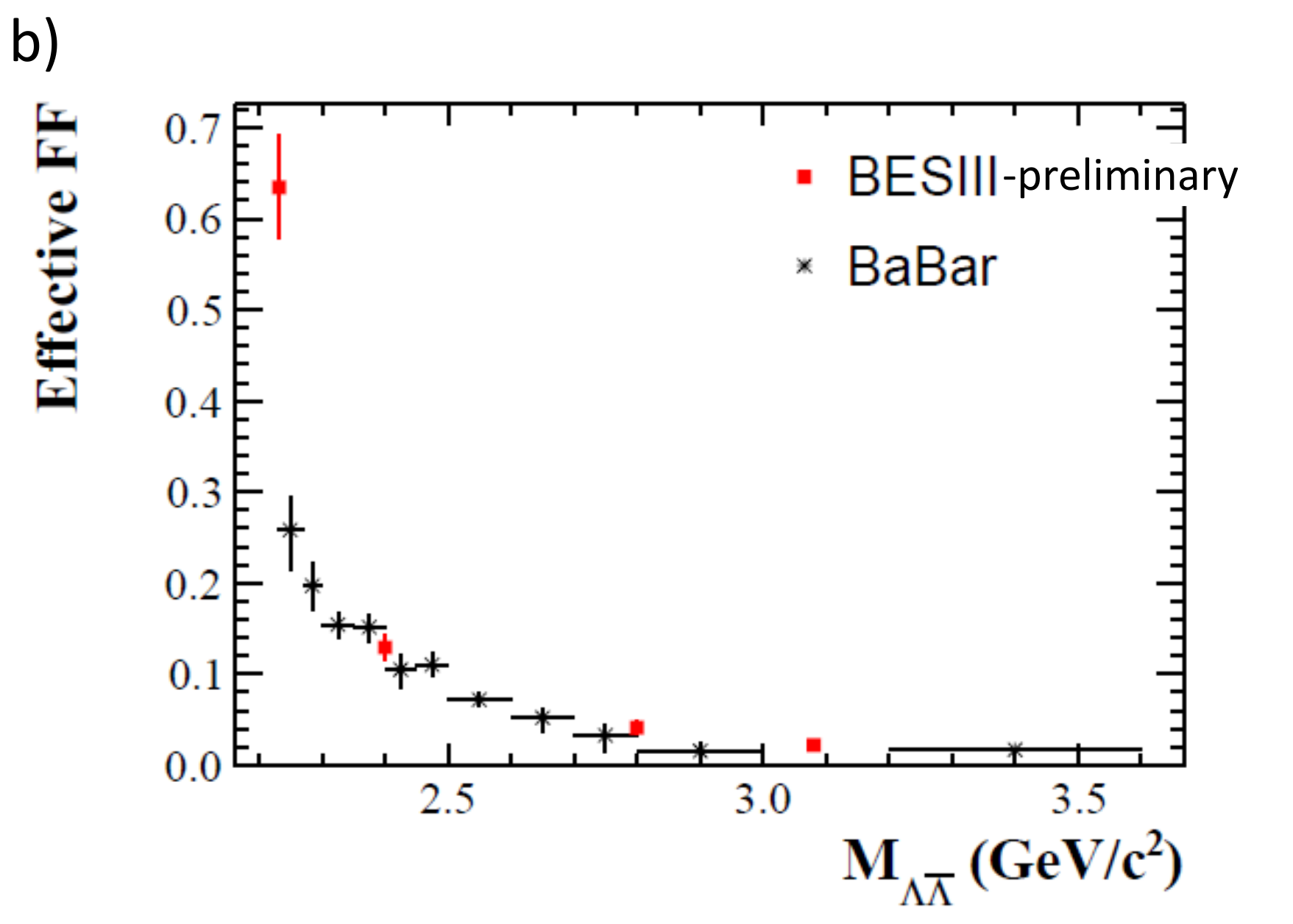}
\end{minipage}
\hspace{\fill}
\caption{\footnotesize {\bf a)} Preliminary BESIII $\ee\rt\lm\lmb $ cross section measurements~\cite{bes_llb} along
with previous results from BaBar~\cite{babar_llb} and DM2~\cite{dm2_llb}.   
{\bf b)} The near-threshold  effective $\Lambda$ form factor inferred from the BESIII and BaBar cross
section measurements (using Eq.\ref{eqn:BBffeff} and ${\mathcal C}=1$).
}
\label{fig:ll-results}
\end{figure}

Preliminary BESIII~\cite{bes_llb} results (including higher $\sqrt{s}$ values measured by more conventional techniques)
are shown in Fig.~\ref{fig:ll-results}a along with published results from BaBar~\cite{babar_llb} and
DM2~\cite{dm2_llb}.  There is good agreement for energies where the different experiments overlap, and the lowest energy
BESIII point shows that the trend of cross sections rising towards threshold, first seen by BaBar, persists
down to $\langle \beta\rangle = 0.03$.  This rise is especially dramatic in the inferred effective form
factor shown in Fig.~\ref{fig:ll-results}b.  The anomalous-looking threshold
behavior may be related to the apparent change in the BaBar-reported $\cos\theta$ distribution from
being nearly isotropic for $\sqrt{s}>2400$~MeV ($\langle \beta \rangle > 0.37$) and more
$\sin^2 \theta$-like for $\sqrt{s}<2400$~MeV ($\langle\beta\rangle <0.37$)~\cite{babar_llb}).  

Corrections to ${\mathcal C}$ and effects of fsi are largely confined to the $S$-wave and
are isotropic.  Therefore, observation of a non-isotropic $\cos\theta$ distribution at threshold
is a more robust indication of non-analytic behavior of the form factor than anomalous cross
section measurements.  Measurements of  $\ppbar$, $\nnbar$ and $\lm\lmb$ angular distributions for
$\langle\beta\rangle\simeq 0.03$ are very difficult at
BESIII, even with more data.  This is not so for $\ee\rt\lm_c^+\bar{\lm}_c^-$;  BESIII has operated
just above the 2$m_{\lm_c}$ threshold and we look forward to results from the analysis of these data.   

\section{Tetraquark mesons?...Pentquark baryons?}

\subsection{The $Z(4430)$ tetraquark candidate}

\begin{figure}[htb]
\begin{minipage}[t]{75mm}
  \includegraphics[height=0.88\textwidth,width=1.0\textwidth]{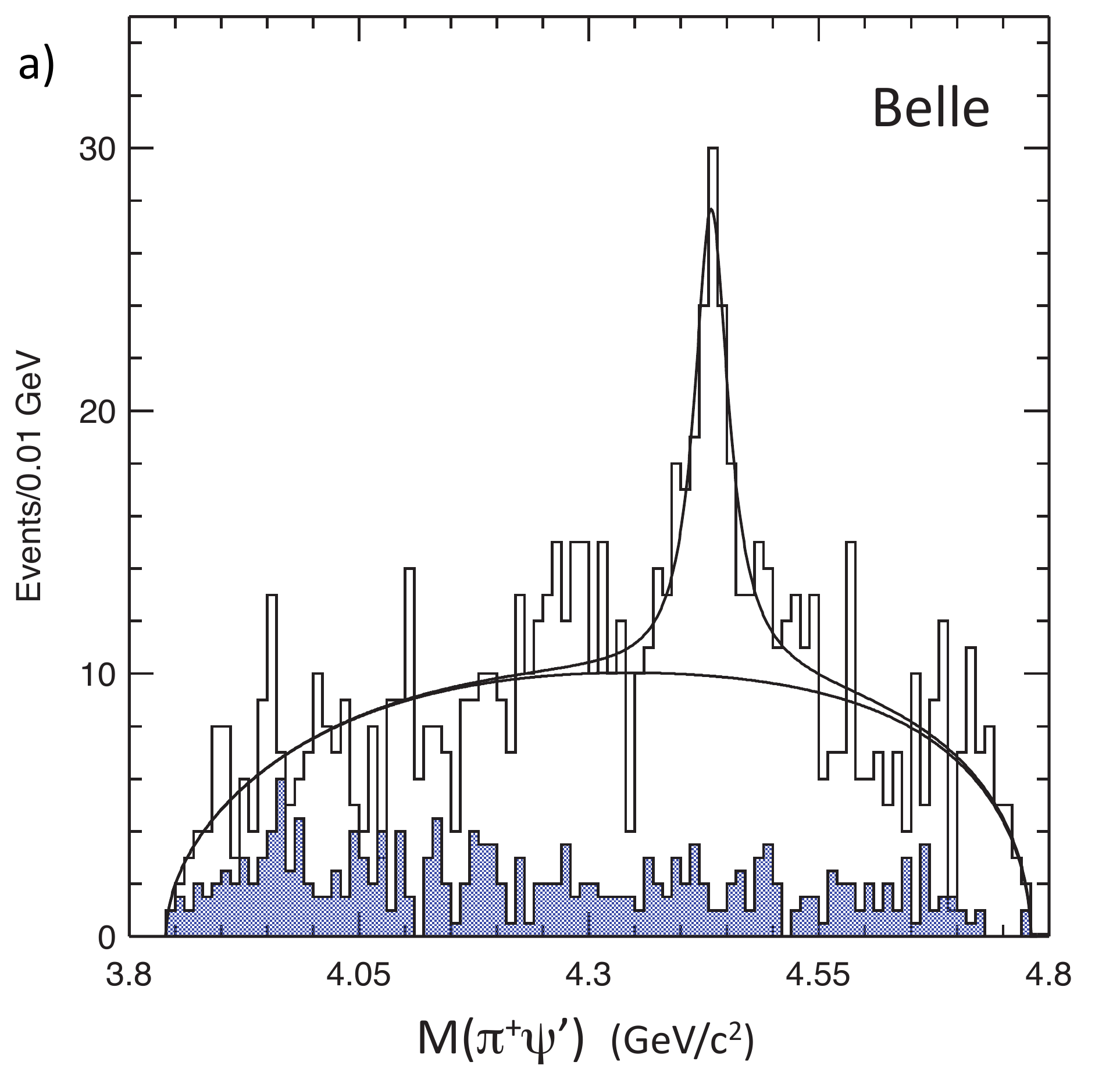}
\end{minipage}
\begin{minipage}[t]{75mm}
  \includegraphics[height=0.88\textwidth,width=1.0\textwidth]{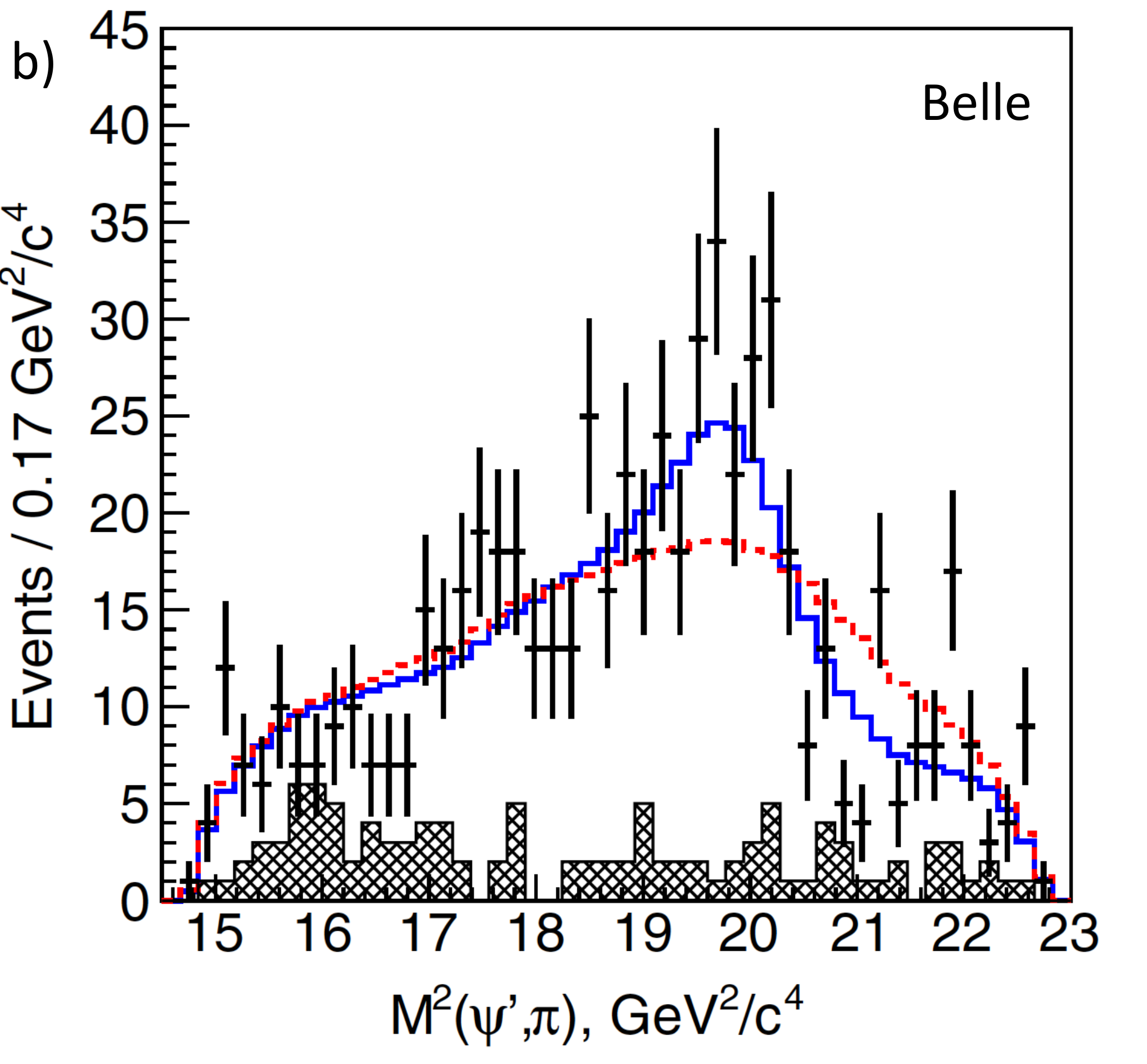}
\end{minipage}
\hspace{\fill}
\caption{\footnotesize
{\bf a)} (Figure~2 from ref.~\cite{belle_z4430}.)
The open histogram shows the $\pip\psip$ invariant mass distribution from $B\rt K\pip\psip$ decays from
Belle~\cite{belle_z4430} for events with the $K^*$ veto requirement applied.  The shaded histogram is
non-$\psip$ background, estimated from $\psip$ mass sidebands.  The curves shows fit results that returned
mass and width values quoted in the text.  
{\bf b)}  (From Fig.~6 in ref.~\cite{belle_z_dalitz2}.)
The data points show the Belle $M^2(\pip\psip)$ distribution with the $K^*$ veto applied.
The solid blue histogram shows a projection of the Belle 4D fit results
with a $Z^+\rt\pip\psip$ resonance included. 
The dashed red
curve shows fit results with no resonance in the $\pip\psip$ channel.  
}
\label{fig:z4430-1}
\end{figure}

Among the tetraquark candidates, I focus on the $Z(4430)$, which has been the most carefully studied.
It was first reported by Belle  in 2007~\cite{belle_z4430} as the $M(\pip\psip)$
mass peak in  $B\rt K\pip\psi'$ decays that is evident in Fig.~\ref{fig:z4430-1}a.  The final state
$\psip$ plus the non-zero electric charge requires a minimal $\ccbar u\bar{d}$ four-quark substructure.
Thus, if the $Z(4430)$ is a genuine resonance, it is a smoking gun signal for a four-quark (tetraquark)
meson.  In the 2007 Belle paper, the influence of the dominant $B\rt K^*(890)\psip$ and $K^*_2(1430)\psip$
decay channels was reduced by excluded events with $K\pi$ invariant masses within $\pm  100$~MeV of the
$K^*(890)$ or $K^*_2(1430)$ masses (the ``$K^*$ veto'').  The distinct peak is fitted with a Breit-Wigner
(BW) resonance on an incoherent background.  The BW signal from the fit has a statistical significance
of $\sim 8\sigma$, with a mass and width of $M=4433\pm 5$~MeV and $\Gamma = 45^{+35}_{-18}$~MeV. 

A BaBar study of the $B\rt K\pip\psip$ decay channel neither confirmed nor contradicted the Belle result~\cite{babar_z4430}.
Although BaBar saw an excess of events in the same $M(\pip\psip)$ region as the Belle signal, their fit using Belle's
mass and width values yielded a statistically marginal ($\sim 2\sigma$)~$Z(4430)\rt\pip\psip$ signal. Belle responded
to concerns about the possibility of ``reflection'' peaks in the $M(\pip\psip)$ distribution, produced by
interference between different partial waves in the $K\pi$ channels, by doing a coherent amplitude analysis
of the $B\rt K\pip\psip$ decay process that used a kinematically complete, four-dimensional (4D)
set of variables: $M(K\pip)$; $M(\pip\psip)$; the $\psip\rt\leplep$ decay helicity angle
and the  angle between the $K\pip$ and $\psip\rt\leplep$ decay planes~\cite{belle_z_dalitz2}. An isobar
model amplitude that included all known $K\pi$ resonances and allowed for contributions from possible
additional, unknown ones, was applied.   This analysis confirmed the existence of a resonance in the $\pip \psip$
channel with greater than $6\sigma$ significance, but with larger mass and width values than those
of the ref.~\cite{belle_z4430} analysis: $M=4485^{+36}_{-25}$~MeV and $\Gamma=200^{+48}_{-58}$~MeV. 

The source of the upward mass and width shifts from the original Belle results can be
seen in Fig.~\ref{fig:z4430-1}b, which shows a comparison of projections of the 4D fit results with
the experimental $M^2(\pip\psip)$ distribution with the $K^*$ veto applied.  The dashed red histogram
shows the best fit results with no resonance in the $\pip\psip$ channel; the solid blue histogram shows
results with the inclusion of a single $\pip\psip$ resonance, where strong interference effects that are
constructive below, and destructive above, the resonance mass, are evident. The ref.~\cite{belle_z4430}
and~\cite{babar_z4430} analyses neglected effects of interference between the $Z\rt \pi \psip$
and the $K^*\rt K\pi$ amplitudes, and only fitted the low-mass lobe
of the dipole-like interference pattern, with a resultant lower mass and narrower width.  

\begin{figure}[htb]
\begin{minipage}[t]{75mm}
  \includegraphics[height=0.78\textwidth,width=1.0\textwidth]{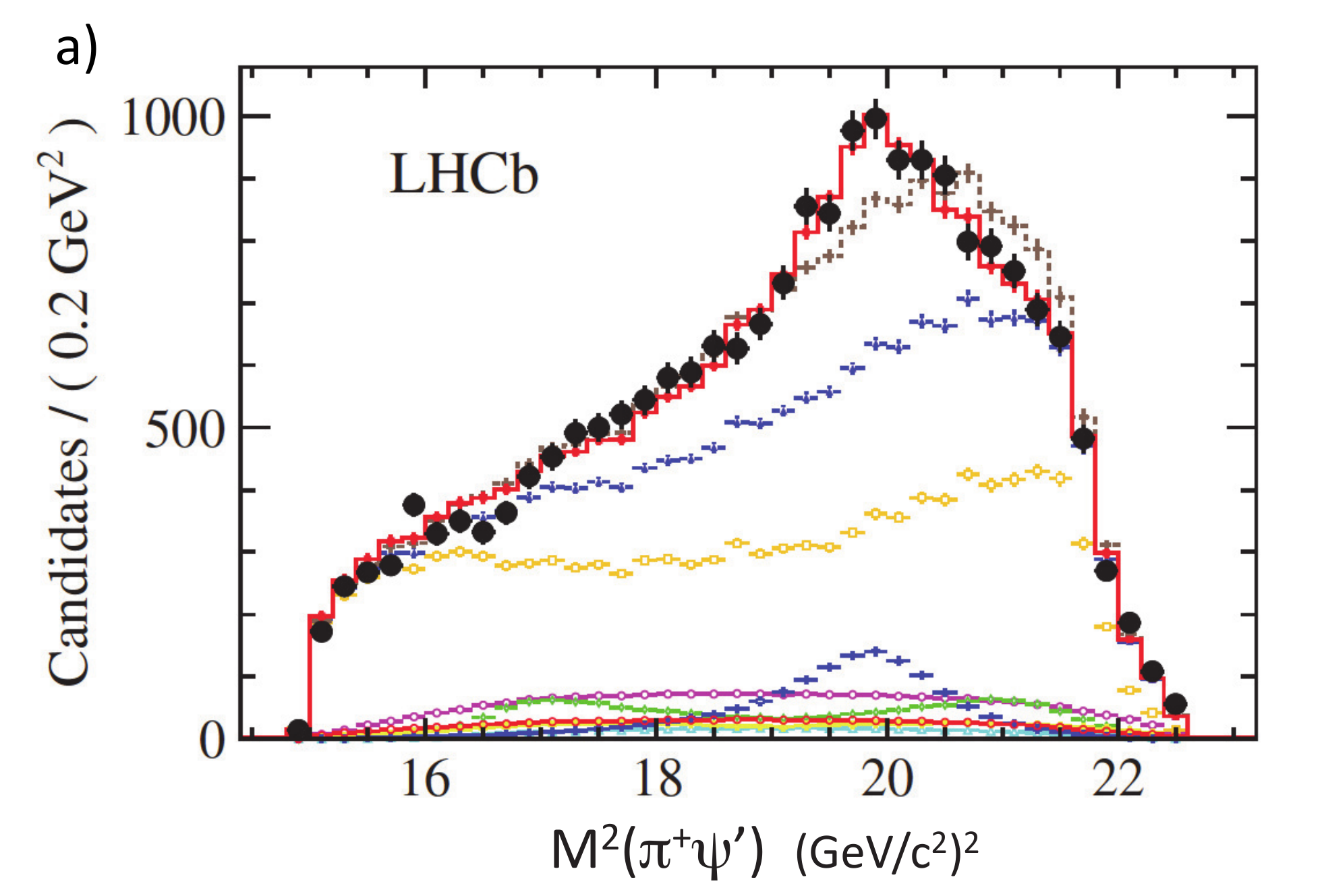}
\end{minipage}
\begin{minipage}[t]{75mm}
  \includegraphics[height=0.73\textwidth,width=0.85\textwidth]{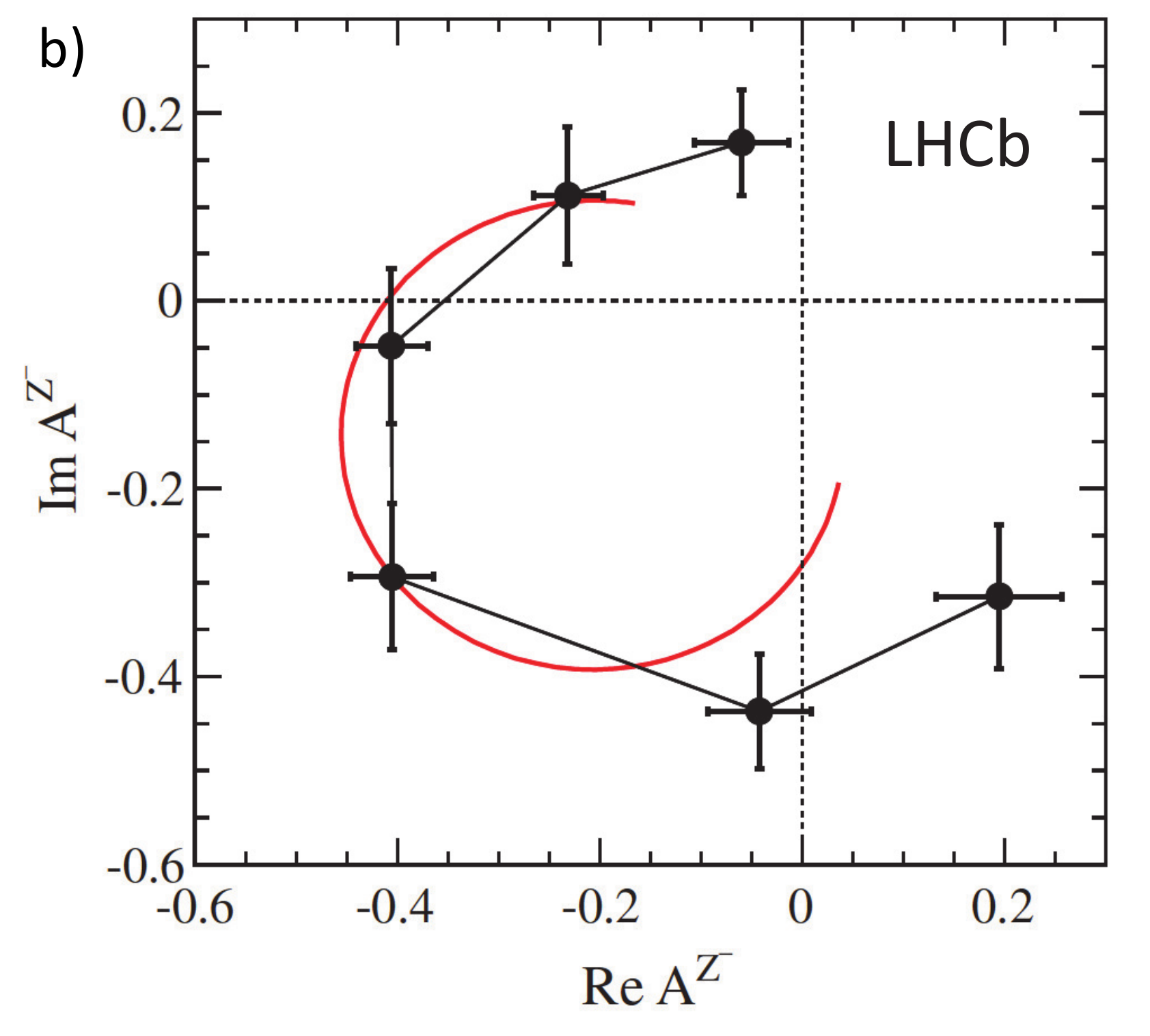}
\end{minipage}
\hspace{\fill}
\caption{\footnotesize  {\bf a)}  (Figure~2 from ref.~\cite{lhcb_z4430}.) 
The LHCb $M^2(\pip\psip)$ distribution for all events (no $K^*$ veto) with
projections from the 4D fits. The solid red histogram shows the fit that includes a 
$Z^+\rt\pip\psip$ resonance term; the dashed brown histogram shows the fit with no resonance in the
$\pip\psip$ channel.
{\bf b)} (Figure~3 from ref.~\cite{lhcb_z4430}.)
The Real (horizontal) and Imaginary (vertical) parts of the ($J^P =1^+$) $Z^+\rt\pip\psip$
amplitude for six mass bins spanning, counter-clockwise, the $4430$~MeV mass region. The
red curve shows expectations for a BW resonance amplitude.
}
\label{fig:z4430-2}
\end{figure}

The big news in 2014 was the confirmation of the Belle $Z(4430)\rt\pip\psip$ claims by the LHCb 
experiment~\cite{lhcb_z4430}, based on a data sample containing $\sim$25k $B^0\rt K^-\pip\psip$
events, an order of magnitude larger than the event samples used by either Belle or BaBar. They find
that their $M(\pip\psip)$ mass distribution cannot be reproduced by reflections from the $K\pi$
channel either with a model-dependent assortment of $K\pi$ resonances up to $J=3$, or by a
model-independent approach that determines Legendre polynomial moments up to fourth order
($J_{K^*}\le 2$) in $\cos\theta_{K^*}$ in bins of $K\pi$ mass, where $\theta_{K^*}$ is the $K\pi$
helicity angle, and reflects them into the $\pip\psip$ channel. The application of a 4D amplitude analysis
that includes a BW resonance amplitude in the $\pip\psip$ channel results in a $Z(4430)$ signal with a huge,
$\sim 14\sigma$,~statistical significance and mass \& width values ($M=4475^{+17}_{-26}$~MeV \&
$\Gamma=172^{+39}_{-36}$~MeV) that are consistent with the Belle 4D analysis results. A comparison of the LHCb
fit results with the data is shown in Fig.~\ref{fig:z4430-2}a,  where strong interference effects, similar
to those seen by Belle (Fig.~\ref{fig:z4430-1}b), are evident.

The LHCb group's large data sample enabled them to relax the assumption of a BW form for the
$Z^+\rt\pip\psip$ amplitude and directly measure the real and imaginary parts of the $1^+$ $\pip\psip$
amplitude in bins of $\pip\psip$ mass.  The results are shown as data points in the Argand plot in
Fig.~\ref{fig:z4430-2}b. There, the phase motion near the resonance peak agrees well with expectations
for a BW amplitude as indicated by the circular red curve superimposed on the plot.  The rapid phase
motion near amplitude-maximum is characteristic of a BW-like resonance.  (The orientation of the red
circle reflects the phase angle between the $B\rt KZ$ and $B\rt K^*(890)\psip$ decay amplitudes.) 

The $Z(4430)$ saga has been a ``game changer'' in the multiquark research community. 
Figures~\ref{fig:z4430-1}b~and~\ref{fig:z4430-2}b clearly show that the signal for a resonance in cases
where there is a large coherent background is not a simple BW-like peak but, instead, a dipole-like shape
with adjacent regions of constructive and destructive interference effects.  This, plus the LHCb Argand
plot shown in Fig.~\ref{fig:z4430-2}b, indicate that future experimental and theoretical studies that do
not include interference and phase behavior will not be considered adequate.

\subsection{The $P_c(4380)$ and $P_c(4450)$ pentaquark candidates}

The expectation that five-quark baryon states, commonly called pentaquarks~\cite{lipkin87}, exist has been
around since Gell-Mann and Zweigs's original quark-model papers, and experimental searches for them have a long,
often controversial, history that is reviewed elsewhere~\cite{roos82,schumacher05,hicks12}. A result of this
history is that experimenters treat this subject with caution.  Thus, when the LHCb group saw signs
of what might be a pentaquark in the $M(\jpsi\ p)$ distribution in $\Lambda_b^0\rt K^- \jpsi\ p$ decays, they
did a very careful and thorough analysis before reporting their results~\cite{lhcb_5q}.
 
\begin{figure}[htb]

\begin{minipage}[t]{75mm}
  \includegraphics[height=0.75\textwidth,width=0.95\textwidth]{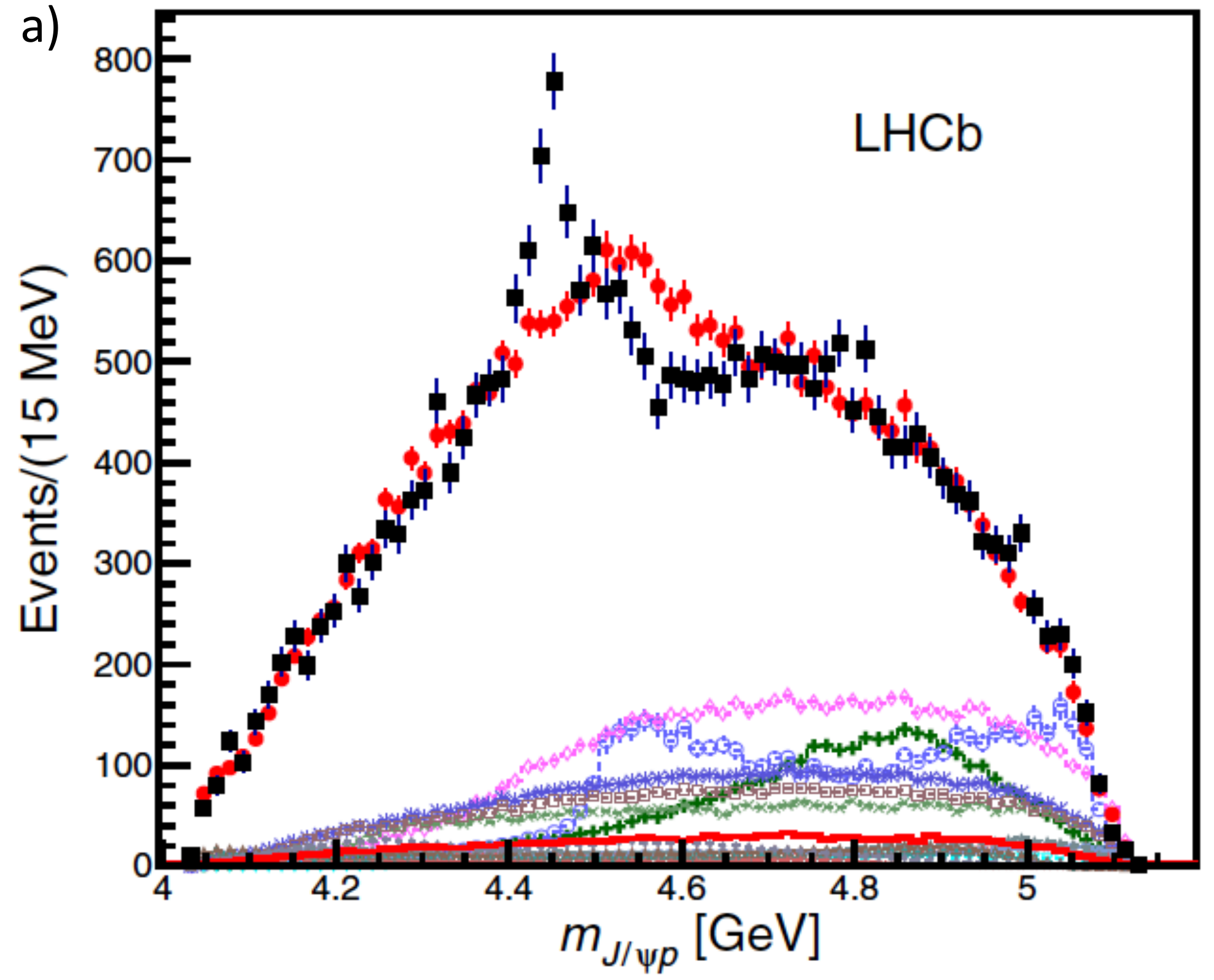}
\end{minipage}
\begin{minipage}[t]{75mm}
  \includegraphics[height=0.75\textwidth,width=0.95\textwidth]{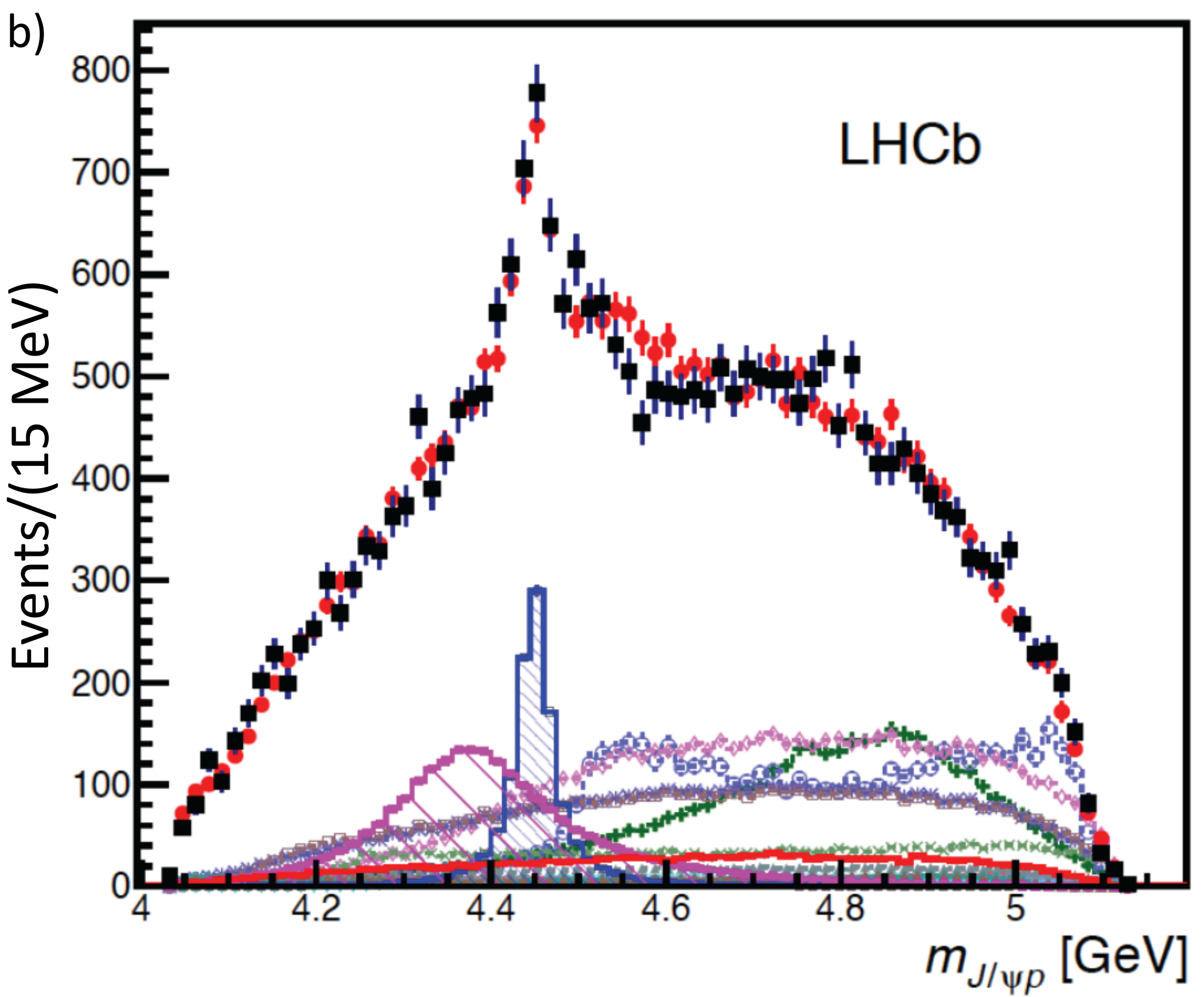}
\end{minipage}
\hspace{\fill}
\caption{\footnotesize  {\bf a)} (Figure~6b from ref.~\cite{lhcb_5q}.)
The solid squares show LHCb's $M(\jpsi\ p)$ distributions and the solid red dots curve show
the results of the fit that includes an extensive set of $\lm^*\rt Kp$ states but no resonances
in the $p\jpsi$ channel. 
{\bf b)} (Figure~8 from ref.~\cite{lhcb_5q}.)
The LHCb $M^2(\jpsi\ p)$ distributions and results of fits that include the $P_c(4380)$
and the $P_c(4450)$ $P_c\rt \jpsi\ p$ resonances described in the text.
}
\label{fig:pc-1}
\end{figure}

The black squares with error bars in Fig.~\ref{fig:pc-1}a show the $M(\jpsi\ p)$ distribution for
a very clean (background $<6$\%) 26k event $\Lambda_b^0\rt K^- \jpsi\ p$ data sample, where there is a striking
peaking structure near $4.4$~GeV.  The LHCb group tried to fit the data with isobar models that
included a number of $K^- p$ resonances with and without resonance terms in the $\jpsi\ p$ channel.
Because of the non-zero $\Lambda_b^0$ spin, the complete kinematic characterization of these decays
requires six independent parameters, for which the LHCb group uses $M(K^-p)$ and five angles.

The red dots with errors~\cite{mc-errors} in Fig.~\ref{fig:pc-1}a show the $M(\jpsi\ p)$ projection of
the best fit that could be achieved using an extensive set of $\Lambda^*\rt K^- p$ resonances and no
resonances in the $\jpsi\ p$ system.  This fit, which reproduces the $M(K^- p)$ distribution well,
undershoots the $M(\jpsi\ p)$ data in the 4.4~GeV peak region and overshoots it at higher masses,
similar to the case for the $Z(4430)$ described in the
previous subsection. The best fit to the data, shown as the red dots in Fig.~\ref{fig:pc-1}b is
accomplished with a model that includes two resonances in the $\jpsi\ p$ channel: a broad $J^{P}=3/2^-$
resonance with $M=4380\pm8\pm29$~MeV and $\Gamma=201\pm 18\pm 86$~MeV, and a narrow $5/2^+$ resonance
with $M=4449.8\pm 1.7\pm 2.5$~MeV and $\Gamma= 39\pm 5 \pm 19$~MeV, where the first (second) error is
statistical (systematic).  The  statistical significance of each state is over $9\sigma$, but the
best fit $(3/2^-,5/2^+)$ assignment is only marginally better than $(3/2^+,5/2^-)$ or $(5/2^+,3/2^-)$,
which cannot be ruled out.  Although there is some ambiguity in the $J^P$ assignments, all the acceptable
fits find that the $P_c(4380)$ and $P_c(4450)$ have opposite parities. 

\begin{figure}[htb]
\begin{center}
\begin{minipage}[t]{80mm}
  \includegraphics[height=0.50\textwidth,width=1.0\textwidth]{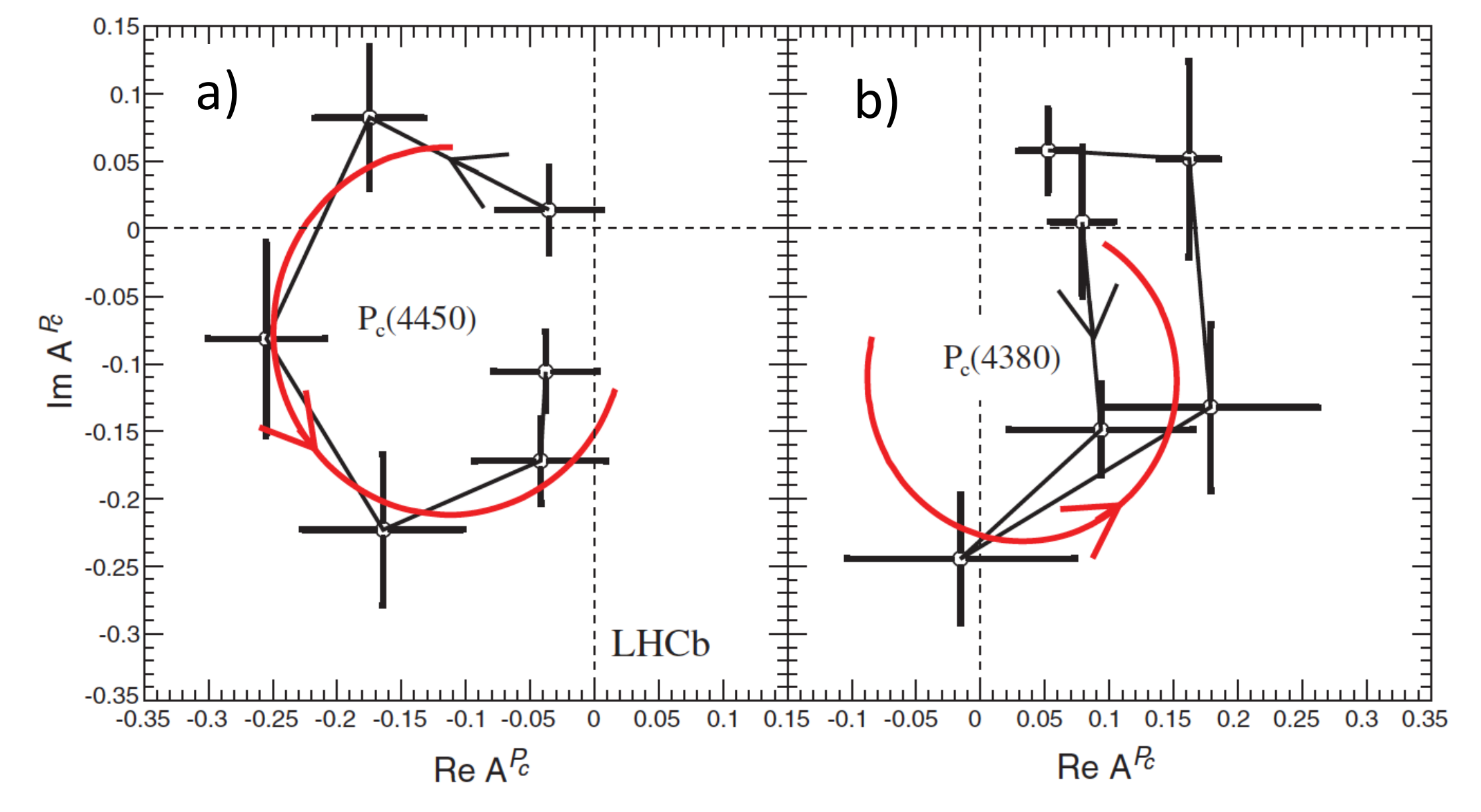}
\end{minipage}
\caption{\footnotesize (Figure~9 from ref.~\cite{lhcb_5q}.)
The Real (horizontal) and Imaginary (vertical) parts of: {\bf a)} the $5/2^+$ $\jpsi\ p$
amplitude for six equal-width mass bins that span, counter-clockwise, the $4410.8 - 4488.8$~GeV
$M(\jpsi\ p)$ region;  {\bf b)} the corresponding $3/2^-$ $\jpsi\ p$ amplitude spanning the $4175 - 4585$~GeV
$M(\jpsi\ p)$ region. 
}
\label{fig:pc-2}
\end{center}
\end{figure}

Figure~\ref{fig:pc-2} shows Argand plots for the two $P_c$ resonant amplitudes.  While both plots show
strong phase motion in the vicinity of the resonance mass peaks, only the higher mass $P_c(4450)$ shows
good agreement with expectations for a BW amplitude. This may be due in part to the broad and less
distinct nature of the the $P_c(4380)$, which makes the shape of this amplitude more sensitive to the
details of the $\Lambda^*$ contributions to the fit.  On the other hand, this may be an indication of
a more complex structure that the statistical accuracy of the current LHCb data sample is unable to
resolve.  

The LHCb results provide strong evidence for the existence of pentaquark systems.  However, their underlying
structure remains unclear.  Cusp models require, and molecule models like, nearby thresholds~\cite{molina10}.
Table~\ref{tbl:1} shows the thresholds for $\bar{D}$ or $\bar{D}^*$ meson and  $\Lambda_c$, $\Sigma_c$ or
$\Sigma_c^*$, and $\chi_{c0}p$ \& $\chi_{c1}$ baryon combinations, where there do not seem to be any compelling
matches for the measured $P_c$ masses of 4380~and~4450~MeV.  The $\Sigma_c^*$-$\Sigma_c$ mass difference is
similar to that for the $P_c(4450)$-$P_c(4380)$, but the $\Sigma_c^*$ and $\Sigma_c$ have the same parity, while
the parities of the $P_c(4450)$ and $P_c(4380)$ are opposite.  The close coincidence of the $P_c(4450)$ mass and
$m_p +m_{\chi_c1}$ prompted a suggestion that the peak may be due to a $p\chi_{c1}$ threshold cusp~\cite{guo15},
but there is no similar threshold coincidence for the $P_c(4380)$. Similarly, some explicit molecular model 
calculations claim success for one, but usually not both, of the
$P_c$ states~\cite{rosner15,zhu15,oset15,he15,meissner15}.
Tightly bound diquark-diquark-antiquark systems have been proposed that account
for both $P_c$ states~\cite{maiani15,lebed15,xghe15,anisovich15,ghosh15,wang15}.

\begin{table}[tbh]
\begin{center}
\caption{Some thresholds in the vicinity of the $P_c(4380)$ and $P_c(4450)$ pentaquarks (in MeV).}
\label{tbl:1}
\begin{tabular}{l|ccc||l|cc}
\hline
~~~~~~~~~~~~& $\Lambda_c(2285)$ & $\Sigma_c(2455)$ & $\Sigma_c^*(2520)$ &      & $\chi_{c0}(3415)$ & $\chi_{c1}(3511)$\\
\hline
$D(1870)$   &    $4155$         &      $4325$      &      $4390$       & $N(940)$ & $4355$       &  $4451$    \\
$D^*(2010)$ &    $4295$         &      $4465$      &       $4530$      &          &              &             \\
\hline
\end{tabular}
\end{center}
\end{table}

\section{Some comments on future possibilities}

The Argand plots for the $Z(4430)$ (Fig.~\ref{fig:z4430-2}b) and 
$P_c(4450)$ (Fig.~\ref{fig:pc-2}a) seem to be pretty convincing evidence that
the $Z(4430)$ and the $P_c(4450)$ are genuine hadron resonances, in which case 
examples of a $\ccbar u\bar{d}$ tetraquark and a $\ccbar uud$ pentaquark have been
established.  What is not so clear is their basic underlying structure.  Are they
extended molecule-like arrangements of color singlet hadrons that are loosely bound
by nuclear-like forces? ...or compact structures containing colored diquarks tightly
bound by the strong color force? ...or some quantum mechanical mixture of these
(and other) possibilities?  Now there are no strong compelling reasons to prefer
one over the other scenarios.  Hopefully, experimental observations of more
states, other decay modes and different production mechanisms may give us some
clues.  Among the other $XYZ$ states, not mentioned in this report, Argand plots
that conclusively show that these are true resonances, and not artifacts of 
kinematic or coupled channel effects~\cite{bugg11,matsuki13,swanson15} would be
welcome.  The above could be done with large data samples collected by the BESIII detector~\cite{besiii_physics}
at energies near the $Y(4260)$ and $Y(4360)$ peaks and, sometime in the future, with
the huge data samples that will be available to BelleII~\cite{belleii_physics}
and the upgraded LHCb detector~\cite{lhcb_upgrade}.

Although a number of multiquark hadron candidates have been reported in the $c$- and $b$-quark
sectors, nothing similar has been forthcoming in the light- and strange-quark sector.
Perhaps this will change with the operation of the strangeness physics program at J-PARC~\cite{jparc}
and, eventually, the PANDA experiment~\cite{panda} at the FAIR facility in GSI.

The peculiarities seen in the baryon time-like form factors discussed in Section~\ref{sec:BBff} may
be hinting at interesting and unanticipated phenomena near thresholds. Moreover, the $X(3872)$,
the first $XYZ$ meson to be reported~\cite{belle_x3872}, has a mass that is
indistinguishable from the $D^0\bar{D}^{*0}$ threshold; the current experimental value
is $(m_{D^0}+m_{\bar{D}^{*0}})-M_{X(3872)}=3\pm 192$~keV~\cite{seth15}. Is this a coincidence? ...or does it 
reflect some dynamics that  have previously been overlooked?  These results suggest
that it may be interesting to systematically investigate the threshold regions of other stable, or nearly stable,
meson-antimeson and baryon-antibaryon pairs. The  BESIII experiment has started to do this for the stable
strange and charmed baryons, and it might be interesting to do similar studies for $D^{(*)}$ and $B^{(*)}$
mesons.  Studies like this may be especially interesting with PANDA, which could explore these
threshold regions with exquisite c.m. energy resolution~\cite{panda}.

\section{Summary}

I think it is safe to conclude that four-quark mesons and five-quark baryons have been
observed.  Their underlying nature is still unknown and considerable experimental
and theoretical work remains to be done before a satisfying understanding of these
states will be achieved.

These new multiquark hadron discoveries have, so far, been confined to the $c$-quark
and $b$-quark sectors; there is evidence for multiquark states in the $s$-quark
or the light quark ({\it i.e.,} $u$- and $d$-quark) sectors but it is less compelling.
Unusual features in the near-threshold, time-like form factors of
the proton, neutron and $\Lambda$ have been seen. While these could be due to an inadequate
understanding of the coulomb factor, a more intriguing possibility is that they are 
due to the influence of baryonium-like states near these thresholds.  

In the field of multiquark mesons and baryons, the game has changed.  Bump hunting
(and bump predicting) is going out of style and more sophisticated analyses that take into account
interference between coherent amplitudes that represent both well known and newly discovered
states are now essential for both theorists and experimentalists.   

\section{Acknowledgements}
I congratulate the organizers of HYP2015 on their interesting and successful meeting, and thank them
for their support.  I thank Sheldon Stone for reading this manuscript and providing valuable
suggestions that improved it. I also acknowledge support from the Institute for Basic Studies (Korea)
under Project Code IBS-R016-D1.


\begin{thebibliography}{9}
\bibitem{mc-errors}  Monte Carlo events generated with the best fit model are used to
produce the projected distributions for comparisons with data.  Thus the fit results
have statistical errors.
\end{thebibliography}
\end{document}